\documentclass[
superscriptaddress,
 amsmath,amssymb,
 aps,
pra,
floatfix,
twocolumn,
]{revtex4-1}

\usepackage{graphicx}
\usepackage{dcolumn}
\usepackage{bm}

\usepackage{amsmath,amsfonts}
\usepackage{amssymb}
\usepackage[colorinlistoftodos]{todonotes}
\usepackage{sistyle}
\usepackage{dsfont}
\usepackage{tikz}
\usepackage{commath}
\usepackage{float}
\usepackage{textcomp}
\usepackage[T1]{fontenc}
\usepackage{fancyhdr}
\usepackage{tabularx}
\usepackage[ pdftex,colorlinks=true,
pdfstartview=FitV,
linkcolor= blue,
urlcolor= blue,
hyperindex=true,
hyperfigures=false]
{hyperref} 

\usepackage{heppennames}

\DeclareMathOperator{\arctanh}{arctanh}

\newcommand{\meanvalue}[1]{\langle#1\rangle}
\newcommand{\ket}[1]{|#1\rangle}
\newcommand{\bra}[1]{\langle#1|}

\newcommand{\sigmajx}{{\hat{\sigma}_j}^x}

\newcommand{\sigmajz}{{\hat{\sigma}_j}^z}

\newcommand{\Jz}{\hat{J}_z}
\newcommand{\Jx}{\hat{J}_x}
\newcommand{\Jy}{\hat{J}_y}
\newcommand{\Jplus}{\hat{J}_+}
\newcommand{\Jminus}{\hat{J}_-}

\newcommand{\Hop}{\hat{H}}
\newcommand{\Sop}{\hat{S}}
\newcommand{\Pop}{\hat{P}}
\newcommand{\Qop}{\hat{Q}}
\newcommand{\Rop}{\hat{R}}

\newcommand{\aop}{\hat{a}}
\newcommand{\adag}{\hat{a}^{\dagger}}
\newcommand{\adagsq}{\hat{a}^{\dagger 2}}

\newcommand{\bop}{\hat{b}}
\newcommand{\bdag}{\hat{b}^{\dagger}}

\newcommand{\bsqr}{\hspace{5pt} \sqrt[]{N-\bdag \hat{b}} \hspace{5pt}}

\newcommand{\dop}{\hat{d}}
\newcommand{\opddag}{\hat{d}^{\dagger}}
\newcommand{\cop}{\hat{c}}
\newcommand{\cdag}{\hat{c}^{\dagger}}
\newcommand{\cdagsq}{\hat{c}^{\dagger 2}}

\newcommand{\modbeta}{|\beta|^2}

\newcommand{\betaconj}{\beta^{\ast}}

\newcommand{\betasqr}{\hspace{5pt} \sqrt[]{N-\modbeta} \hspace{5pt}}

\newcommand{\gprime}{g_{\beta}}
\newcommand{\goneprime}{g_1^{\beta}}
\newcommand{\gtwoprime}{g_2^{\beta}}

\newcommand{\Ko}{\hat{K}_0}
\newcommand{\Kp}{\hat{K}_+}
\newcommand{\Km}{\hat{K}_-}

\newcommand{\Kop}{\hat{K}_0\sp{\prime}}
\newcommand{\Kpp}{\hat{K}_+\sp{\prime}}
\newcommand{\Kmp}{\hat{K}_-\sp{\prime}}

\newcommand{\Uop}{\hat{U}}

\begin{document}

\preprint{APS/123-QED}

\title{Superradiant phase transition in the ultrastrong coupling regime\\ of the two-photon Dicke model}

\author{L. Garbe}
 \email{louis.garbe@ens-lyon.fr}
\affiliation{%
 Laboratoire Mat\'eriaux et Ph\'enom\`enes Quantiques,
Universit\'e Paris Diderot, CNRS UMR 7162, Sorbonne Paris Cit\'e,
France}%
\author{I. L. Egusquiza}
\affiliation{Department of Theoretical Physics and History of Science,
University of the Basque Country UPV/EHU, Apartado 644, E-48080 Bilbao, Spain}

\author{E. Solano}
\affiliation{Department of Physical Chemistry, University of the Basque Country UPV/EHU, Apartado 644, E-48080 Bilbao, Spain}
\affiliation{IKERBASQUE, Basque Foundation for Science, Maria Diaz de Haro 3, 48013 Bilbao, Spain}

\author{C. Ciuti} 
\author{T. Coudreau}
\author{P. Milman}
\author{S. Felicetti}%
\affiliation{%
 Laboratoire Mat\'eriaux et Ph\'enom\`enes Quantiques,
Universit\'e Paris Diderot, CNRS UMR 7162, Sorbonne Paris Cit\'e,
France}%

\date{\today}

\begin{abstract} 
The controllability of current quantum technologies allows to implement spin-boson models where two-photon couplings are the dominating terms of light-matter interaction. In this case, when the coupling strength becomes comparable with the characteristic frequencies, a spectral collapse can take place, {\it i.e.} the discrete system spectrum can collapse into a continuous band. Here, we analyze the thermodynamic limit of the two-photon Dicke model, which describes the interaction of an ensemble of qubits with a single bosonic mode. We find that there exists a parameter regime where two-photon interactions induce a superradiant phase transition, before the spectral collapse occurs. 
Furthermore, we extend the mean-field analysis by considering second-order quantum fluctuations terms, in order to analyze the low-energy spectrum and compare the critical behavior with the one-photon case.

\end{abstract}

\maketitle

\section{Introduction}

In quantum optics, the superradiant phase transition~\cite{Mallory69,Hepp75} (SPT) is the abrupt change in the behavior of the ground state properties of a quantum many-body system, while a physical parameter is continuously varied. It is a quantum phase transition, i.e. it can be accessed at zero temperature and it is due to quantum fluctuations.
The archetypal model known to display such a transition is the Dicke model~\cite{Gelfand54,Brandes05} (DM), which describes the interaction of an ensemble of two-level quantum systems, or qubits, with a single bosonic mode. In the limit of a large number of qubits, the DM undergoes a SPT~\cite{Emary03} in the ultrastrong coupling (USC) regime, where the collective light-matter coupling becomes comparable to the qubit and field bare frequencies~\cite{Ciuti05}. Although the DM is commonly used to describe atomic and solid-state systems, whether it provides a reliable description of the system ground state when approaching the critical coupling is still the subject of debate~\cite{Rzazewski75,Knight78,Gawedzki81,Keeling07,Deliberato14,Vukics14,Griesser16}. 
In particular, the presence of the so-called diamagnatic term is expected to prevent the SPT. The debate has been recently extended to the framework of circuit QED~\cite{Chen07,Lambert09,Nataf10,Viehmann11,Jaako16,Bamba16}, where the USC regime has been experimentally achieved~\cite{Niemczyk10,Forn-Diaz10,Forn-Diaz16,Chen16,Yoshihara16}.

However, a compelling way to circumvent no-go theorems consists in using driven systems to engineer effective Hamiltonians. Indeed, the SPT has been observed in driven atomic systems which effectively reproduce the DM~\cite{Baumann10,Baden14,Klinder15}. In general, driven atomic or solid state systems represent a powerful tool to access the USC regime of quantum optical models, both in few~\cite{Crespi12,Ballester12,Pedernales2015,Langford16,Felicetti16} and many-body physics~\cite{Dimer07,Carusotto13}. In the USC regime, even apparently simple models entail a very complex physics. This is the case for the quantum Rabi model~\cite{Braak11,Rossatto16}, which corresponds to a single-qubit DM. Furthermore, the engineering of effective Hamiltonians allows to implement generalized quantum optical models~\cite{Genway14, Zou14}, including  anisotropic couplings or two-photon interactions. 
 In the case of anisotropic couplings, reaching the USC regime leads to  parity-symmetry breaking~\cite{Xie14,Cui15} and to a rich phase diagram in the many-body limit~\cite{Baksic14}.

Similarly, the two-photon Rabi model has highly counter-intuitive spectral and dynamical features. Its spectrum collapses into a continuous band for a specific value of the coupling strength~\cite{Emary02,Dolya09,Travenec12,Peng16}. In the transition from the strong to the USC regime of the two-photon Rabi model, a continuous symmetry breaks down into a four-folded discrete symmetry, identified by a generalized-parity operator~\cite{Felicetti15}. However, so far there are no known results on the ground state of two-photon models in the many-body limit.

In this work, we perform first a mean-field analysis of the two-photon Dicke model and we find that the system exhibits a phase transition in the thermodynamic limit. 
This transition is superradiant in the sense that it is characterized by a macroscopic change in the average photon number. 
The boundary of the phase transition is set by the critical value of the coupling strength, which depends on the qubit and field energies. Interestingly, the coupling strength for which the spectrum collapses depends only on the field frequency. For larger values of the coupling strength, the Hamiltonian is not bounded from below and the model is not well defined. We define the parameter regime where the SPT could be accessed within the validity region of the model, that is, where the critical coupling is smaller than the collapse coupling strength. Finally, we go beyond mean-field by including second-order quantum fluctuations. This lets us characterize the system phases and analyze the differences with the SPT of the standard DM. We find fundamental differences in the critical scaling of the bosonic field.

The two-photon DM could be implemented using trapped ions~\cite{Felicetti15},
which have been used to realize spin systems composed of hundreds of qubits~\cite{Britton12}. Similar schemes can be conceived for other atomic or solid state systems. Particularly promising are superconducting devices, where  bosonic modes have been coupled to increasingly large spin ensembles~\cite{Macha14,Kakuyanagi16}. In any implementation, the critical issue would be the number of qubits that can be effectively coupled to a single bosonic mode. In our case, we have considered the thermodynamic limit $N \gg 1$, with $N$ the number of qubits. We show that, in this limit, the  superradiant phase transition of the two-photon DM entails squeezing and spin-squeezing properties. These features provide a signature of the phase transition that could be observed~\cite{Sorensen99,Vitagliano14,Saideh16} for a smaller number of qubits.

\section{\label{sec:level1}Mean-Field Analysis}
\label{mean_field}

\begin{figure}[H]
\begin{center}
\includegraphics[width=0.5\textwidth]{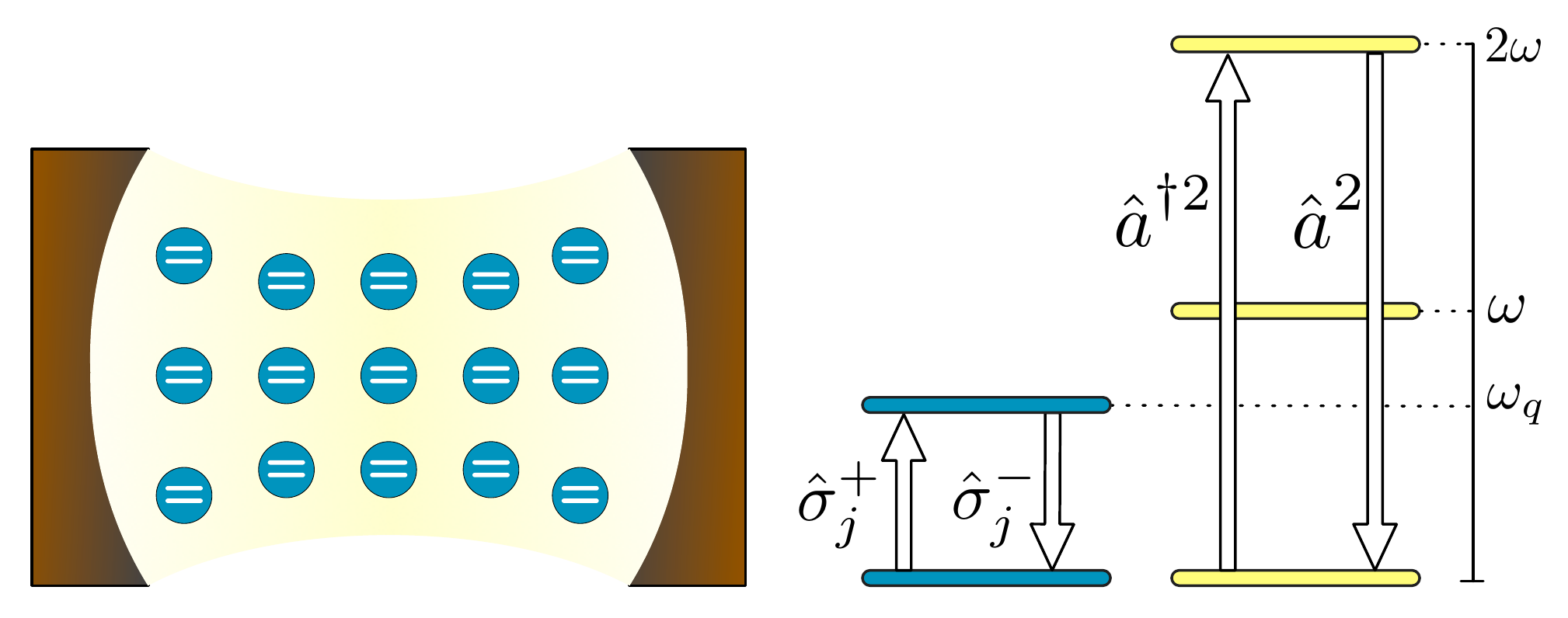}
\caption{(Left) Sketch of an ensemble of qubits interacting with a single bosonic mode via two-photon coupling terms, as in Eq.\eqref{H départ}. (Right) Energy levels of the uncoupled systems. Arrows represent rotating and counter-rotating two-photon transitions. \label{sketch}}
\end{center}
\end{figure}

We consider an ensemble of \textit{N} qubits interacting with a bosonic mode via two-photon interaction, as sketched in Fig.\ref{sketch}. The system Hamiltonian is given by:
\begin{equation}
\hat{H}=\omega \adag \aop + \frac{\omega_q}{2}\sum_{j=1}^N\sigmajz + \frac{g}{N}\sum_{j=1}^N\sigmajx(\aop^2+\adagsq),
\label{H départ}
\end{equation}
with $\hbar=1$, $\sigmajz$ and $\sigmajx$ are Pauli operators describing the j-th ion, and $\aop$ and $\adag$ are bosonic ladder operators. In Ref.\cite{Felicetti15}, Felicetti \textit{et al.} have proposed to implement this model using a chain of ions in a trap illuminated by two lasers, with the motional degree of freedom of the chain playing the role of the bosonic field. The coupling strength $g$, qubit energy spacing $\omega_q$ and bosonic frequency $\omega$ are then effective and tunable parameters which depend on the frequencies and amplitudes of the two lasers. In principle, it is then be possible to reach the ultrastrong coupling (USC) regime: $g \sim \omega$. In the following, we will consider this regime of parameters, as well as the thermodynamic limit $N \gg 1$. 
In the USC regime, the two-photon Dicke model exhibits a spectral collapse \cite{Travenec12, Felicetti15}:  
 for $g=\frac{\omega}{2}$, the energy levels of the system collapse into a continuum. Beyond this limit, the ground state of (\ref{H départ}) is no longer defined, which renders the notion of phase transition meaningless. Thus, the goal of this work is to study the existence of a phase transition for $g<\frac{\omega}{2}$. For this purpose, it is convenient to describe the ensemble of qubits by pseudospin operators: we define  $\Jz=\frac{1}{2}\sum_{j}\sigmajz$, $\Jx=\frac{1}{2}\sum_{j}\sigmajx$, $\hat{J}_{\pm}=\sum_{j}{\hat{\sigma}_j}^{\pm}$ 
, which gives us:
\begin{equation}
\Hop=\omega \adag \aop + \omega_q \Jz + \frac{g}{N}(\Jplus + \Jminus)(\aop^2 + \adagsq).
\label{model avec J}
\end{equation}

In the following, we will speak only in terms of the fluctuations and polarization of this "spin", which are physically interpreted in terms of population and coherence of the states of the qubits. We study the phase diagram of our model by using a mean-field approach, inspired by the analysis performed in \cite{Emary03} and \cite{Baksic14}. The point here is to determine how the properties of the ground state evolve when $g$ increases. First of all, we use the Holstein-Primakoff (HP) transformation to turn our pseudospin operators into bosonic operators:
\begin{equation}
\Jplus=\bdag \bsqr,\hspace{5pt} \Jminus=\bsqr \bop,\hspace{5pt} \Jz=\bdag \bop - \frac{N}{2},
\label{def Hosltein-Primakoff}
\end{equation}
with $[\bop, \bdag]=1$. Here and in the following, we restrict ourselves to the eigenspace of $\overrightarrow{J}^2$ 
associated with eigenvalue $\frac{N}{2}(\frac{N}{2}+1)$ (that is, the maximal angular momentum eigenspace $\mathcal{H}_{\frac{N}{2}}$) \cite{note subspace}.

Next, we shift $\bop$ and $\bdag$ with respect to their mean values in the ground state $\ket{GS}$ of (\ref{model avec J}):   
\begin{equation}
\label{defdop}
\bop=\beta + \dop, \hspace{20pt} \beta=\bra{GS}\bop\ket{GS}, \hspace{20pt} [\dop, \opddag]=1. 
\end{equation}
As a zeroth order approximation, we neglect altogether the spin fluctuations $\dop$ and $\opddag$, which gives:
\begin{subequations}
\begin{equation}
\Hop = \omega\adag \aop + \gprime(\aop^2 + \adagsq) + \omega_q\modbeta - \frac{\omega_qN}{2},
\end{equation}
\begin{equation}
\gprime=\frac{g}{N}(\beta + \betaconj)\betasqr.
\end{equation}
\label{Heff a only}
\end{subequations}
In other terms, we replace the spin operators by their classical mean values. This Hamiltonian is quadratic in $\aop$, and can therefore be diagonalized by Bogoliubov transformation. The ground state is a squeezed vacuum state with squeezing parameter:
\begin{equation}
r_a^{(MF)} =\frac{1}{2}\arctanh(\frac{2\gprime}{\omega}).
\label{def ra mean-field}
\end{equation}
The corresponding ground state energy is given by:
\begin{equation}
E_G = \frac{1}{2}\frac{\cosh(2r_a^{(MF)})}{\omega}(\omega^2 - 4\gprime^2) +\omega_q|\beta|^2 - \omega_q\frac{N}{2} - \frac{\omega}{2}.
\end{equation}
The final step consists in minimizing this energy in order to determine the value of $\beta$ the system adopts in its ground state. $\beta$ then plays the role of order parameter for our system: a change in its behavior leads to a modification of the qualitative properties of the ground state, which indicates a phase transition. We find that, for $g<g_t=\sqrt{\frac{\omega\omega_qN}{4}}$, $E_G$ is minimal for $\beta=\betaconj=0$. For $g>g_t$, we have two degenerate minima: 
\begin{equation}
\beta=\betaconj= \pm \sqrt{\frac{N}{2}} \left(1-\sqrt{\frac{1-\mu}{4\mu^2\lambda^2-\mu}}\right)^{1/2}= \pm \beta_0,
\end{equation}
where we have defined:
\begin{equation}
\lambda=\frac{\omega}{2\omega_qN} \geq 0 \hspace{5pt} ; \hspace{5pt} \mu = \frac{4g^2}{\omega^2} \geq 0.
\end{equation}

Thus, our system exhibits two phases: in the first, $\beta$ and the squeezing parameter are zero for all $g$, meaning that the field $\aop$ is in the vacuum state and the pseudospin $\Jz$ is polarized along the $z$-axis: $\meanvalue{\Jz}=\lvert\beta\rvert^2-\frac{N}{2} = -\frac{N}{2}$, $\meanvalue{\Jx}=\meanvalue{\Jy}=0$. In the second phase, the ground state is twice degenerate. This degeneracy comes from the fact that $\beta$ can be positive or negative. The field $\aop$ is in a squeezed vacuum state, the direction of squeezing depending on the sign of $\beta$:
(i) For $\beta>0$, the quadrature $\hat{X}_a=\aop+\adag$ is squeezed, the fluctuations of $\hat{P}_a=i(\adag-\aop)$ are amplified. (ii) For $\beta<0$, the squeezed quadrature is $\hat{P}_a$.
In both cases, when $g$ increases, the squeezing parameter increases and the pseudospin polarization evolves towards the $x$-axis: $\lvert\meanvalue{\Jz}\rvert$ decreases until it hits zero for $g=g_{collapse}=\frac{\omega}{2}$, and $\lvert\meanvalue{\Jx}\rvert$ increases at the same time. This phase diagram is reminiscent of the one-photon Dicke model and its superradiant phase transition \cite{Emary03,Baksic14}. Here, only the pseudospin acquires macroscopic mean value in the second phase; the mean value of $\aop$ remains zero. However, the average number of photons becomes nonzero at the transition; thus, we argue that our transition may still be dubbed "superradiant".
Note also that the zero mean value of the bosonic field comes from the fact that we have considered only quadratic terms for $\aop$ in the Hamiltonian. If a linear term is present, $\aop$ will acquire nonzero mean value at the mean-field level (See Appendix \ref{appendix linear expansion of H}).

Let us finally notice that, for the phase transition to occur before the spectral collapse, the following condition must be satisfied: $\omega_q N <\omega$. Since $\omega$ and $\omega_q$ are adjustable effective parameters, it is possible in principle to meet this condition. From this point on, we will consider that the order of magnitude of the parameter $\omega_q$ is $\frac{\omega}{N}$ ($\omega_q=O(\frac{\omega}{N})$). We can notice that the approximation that we have made to obtain (\ref{Heff a only}) amounts simply to keeping only terms of order $O(\omega)$ in the Hamiltonian, and to neglecting higher-order terms \cite{note zero omegaq}.

\section{\label{beyond MF}Beyond Mean-Field}
\label{beyond}
Let us now take the fluctuations of $\Jz$ into account, in addition to the mean value. We will expand the Hamiltonian keeping terms of order $O(\omega)$, $O(\frac{\omega}{\sqrt{N}})$ and $O(\frac{\omega}{N})$. Then, we will make use of a technique that Hwang \& \textit{al.} applied to the Rabi and Jaynes-Cumming models \cite{article Hwang JC,article Hwang Rabi}. 
This method is inspired by the Schrieffer-Wolff transformation \cite{ref Schrieffer-Wolff, ref adiabatic elimination}, and can be summarized as follows: one starts with an Hamiltonian describing a spin operator $\hat{\sigma}_z$ and another degree of freedom, that can be written as
\begin{equation}
\Hop = \hat{H}_0 + \epsilon \hat{H}_c,
\label{forme_generale_H}
\end{equation}
with a small parameter $\epsilon \ll 1$, $\hat{H}_0$ an operator that does not couple the $\hat{\sigma}_z$ eigenstates, and $\hat{H}_c$ that does.
In \cite{article Hwang JC} for instance, this method was applied to a Jaynes-Cummings Hamiltonian: 
\begin{equation}
\Hop_{JC}=\omega_0\adag \aop + \frac{\Omega_{spin}}{2}\hat {\sigma}_z - \epsilon \hspace{2pt}g(\aop\hat{\sigma}_{+}+ \adag \hat{\sigma}_-).
\label{H JC Hwang}
\end{equation}

The method consists in decoupling the spin eigenspaces up to a certain order of $\epsilon$. This is done by finding a transformation $e^{\Sop}$, with $\Sop$ an  anti-hermitian operator such that $\Hop\sp{\prime}=e^{-\Sop} \Hop e^{\Sop}$ commutes with $\hat{\sigma}_z$ up to a certain order in $\epsilon$. It is then possible to project $\Hop\sp{\prime}$ in one of the spin eigenspaces, which allows to effectively suppress the spin degree of freedom and to diagonalize the Hamiltonian more easily. 

In our case, it would seem natural to decouple the eigenspaces of the pseudospin $\Jz$. Instead, however, we are going to apply the HP transformation once more, and shift the operators $\bop$ with respect to their mean-field expectation value $\beta$. 
\begin{figure}[h]
\begin{center}
\includegraphics[width=0.75\linewidth]{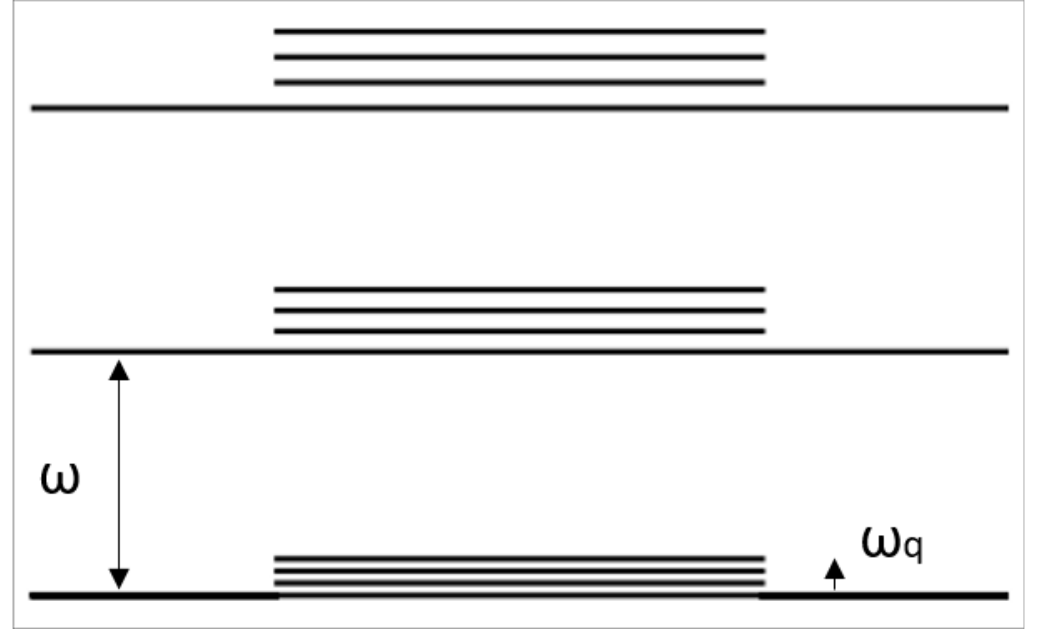}
\end{center}
\caption{Schematic layout of the energy levels of our Hamiltonian in the regime $\omega_q\ll \omega$. In this regime, the energy scale associated to the pseudo-spin operator $\Ko$ defined in Eq.\eqref{def_Ko} is much larger than the energy scale of the bosonic operator $\bop$ of Eq.\eqref{def Hosltein-Primakoff}. \label{energyLevSketch}}
\end{figure}

We define the following operators:
\begin{eqnarray}
\label{def_Ko}
\Ko=\frac{1}{2}(\adag \aop + \frac{1}{2}),\\
\label{def_Kp}
\Kp=\frac{1}{2} \adagsq,\\
\label{def_Km}
\Km=\frac{1}{2} \aop^2,
\end{eqnarray}
which obey spin-like commutation relations $[\Ko,\hat{K}_{\pm}]=\pm \hat{K}_{\pm}$, and $[\Kp, \Km]=-2\Ko$. Note that the commutation algebra here is SU(1,1) instead of SU(2), hence these are not spin operators even if the commutation relations are the same. 
 In order  to apply the method described above, we take profit of these commutation relations to decouple the eigenspaces of these pseudospin operators 
up to a certain order of the small parameter $\frac{1}{\sqrt{N}}$. Then, we project out the $\hat{K}$ degree of freedom. This gives us an effective Hamiltonian describing the low-energy fluctuations of $\bop$ above the ground state; the detailed calculations can be found in Appendix \ref{appendix method Hwang description}, \ref{appendix first phase} and \ref{appendix second phase}.

\begin{figure}[t]
    \centering
	\includegraphics[scale=0.4]{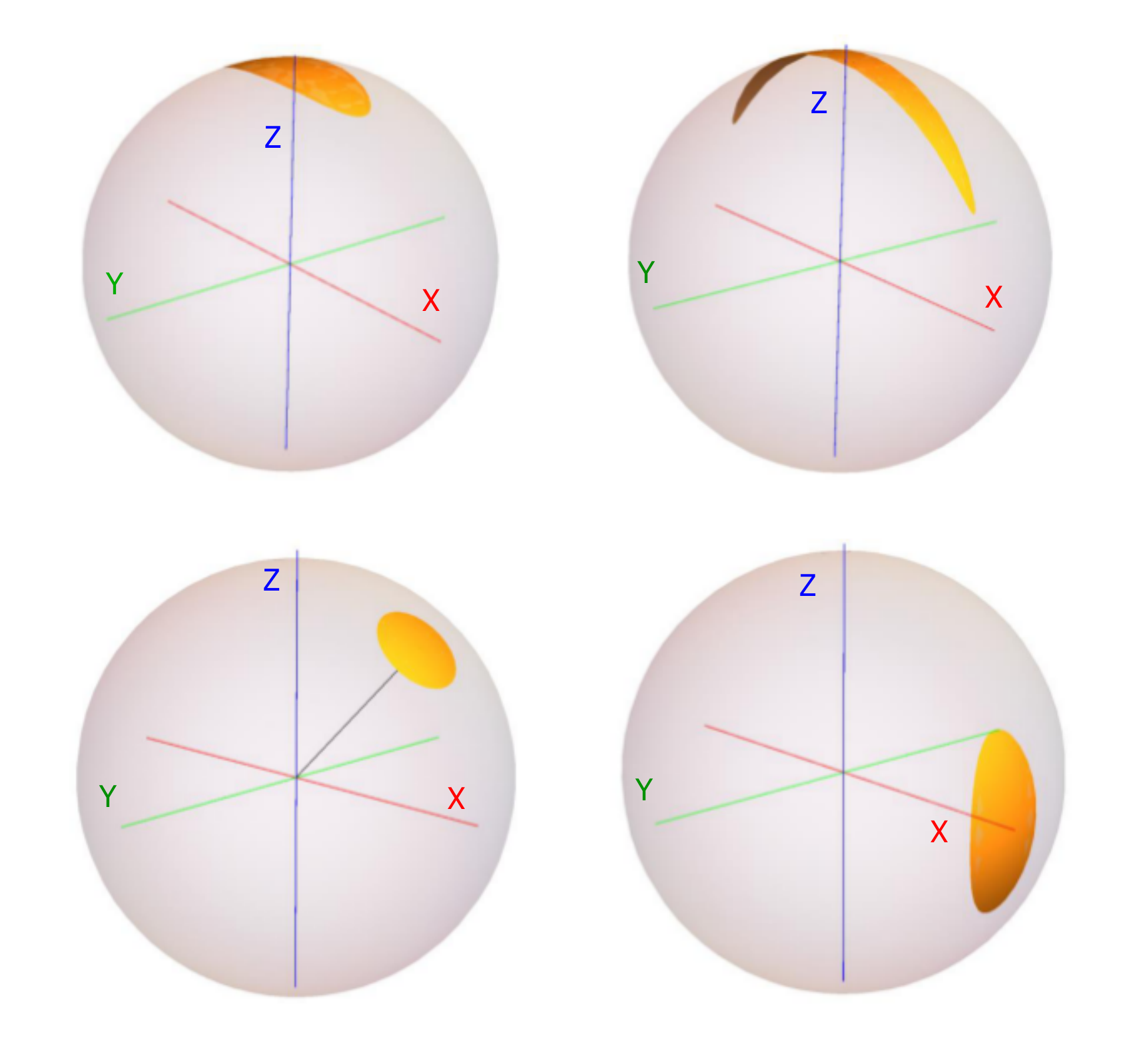}
    \caption{Schematic representation of mean value and spin-squeezing fluctuations of the collective angular momentum $J$ state, in a generalized Bloch sphere. Several values of $g$ are considered, (top left $g < g_t$, top right $g \lesssim g_t$, bottom left $g_t < g < g_{collapse}$, bottom right $g \lesssim g_{collapse}$), for $\beta > 0$. For convenience, the $z$-axis was inverted.}
\label{Spin Bloch sphere}
\end{figure}

While not very intuitive at first glance, this manipulation can be justified by two arguments. Firstly, it is interesting to use the HP transformation on $\Jz$ again in order to make a link to our previous study and to re-use our results. Hence, only $\aop$ is available to play the role of the pseudospin whose eigenspaces we seek to decouple. Next, as we mentioned previously, we consider the following regime of parameters: $\omega_q=O(\frac{\omega}{N})$. Since $N \gg 1$, this means that the typical energy separation between the $\aop$ eigenstates, given by $\omega$, is much bigger than the typical energy separation for the $\Jz$ (or $\bop^\dagger\bop$) eigenstates, given by $\omega_q$. As a consequence, the eigenspaces associated with the pseudospin (\ref{def_Ko}) are well separated in energy, but the eigenspaces of $\Jz$ are not. A schematic representation of the energy levels is shown in Fig.\ref{energyLevSketch}. Thus, it is possible to project the Hamiltonian into the lowest-energy eigenspace of $\Ko$ and to study the fluctuations of $\Jz$ while staying inside this subspace, but not the opposite.

In the first phase, after having decoupled the $\Ko$ eigenspaces and projected the system into the lowest-energy one, we obtain the following Hamiltonian:
\begin{equation}
\Hop = -\frac{\omega_qN}{2} + \omega_q \opddag \dop - \frac{g^2}{N\omega}(\dop+\opddag)^2 + O\left(\frac{\omega}{N\sqrt{N}}\right),
\label{Heff_1erephase}
\end{equation}
where $\dop$ has been defined in Eq.\eqref{defdop}.
The associated ground state for $\dop$ is a squeezed vacuum state, with squeezing parameter $r_s^{(1)}=\frac{1}{4}\ln(1-\frac{4g^2}{N\omega\omega_q})$ (let us note that this parameter is negative, meaning that the squeezed quadrature is $\hat{P}_d$ instead of $\hat{X}_d$). This constitutes a piece of information about the fluctuations of $\Jz$, while only mean values were accessible in our first analysis. The $\aop$ field is found to be in a coherent vacuum state, with no modifications with respect to the mean-field analysis.

\begin{figure}[t]
\begin{center}
\includegraphics[scale=0.57]{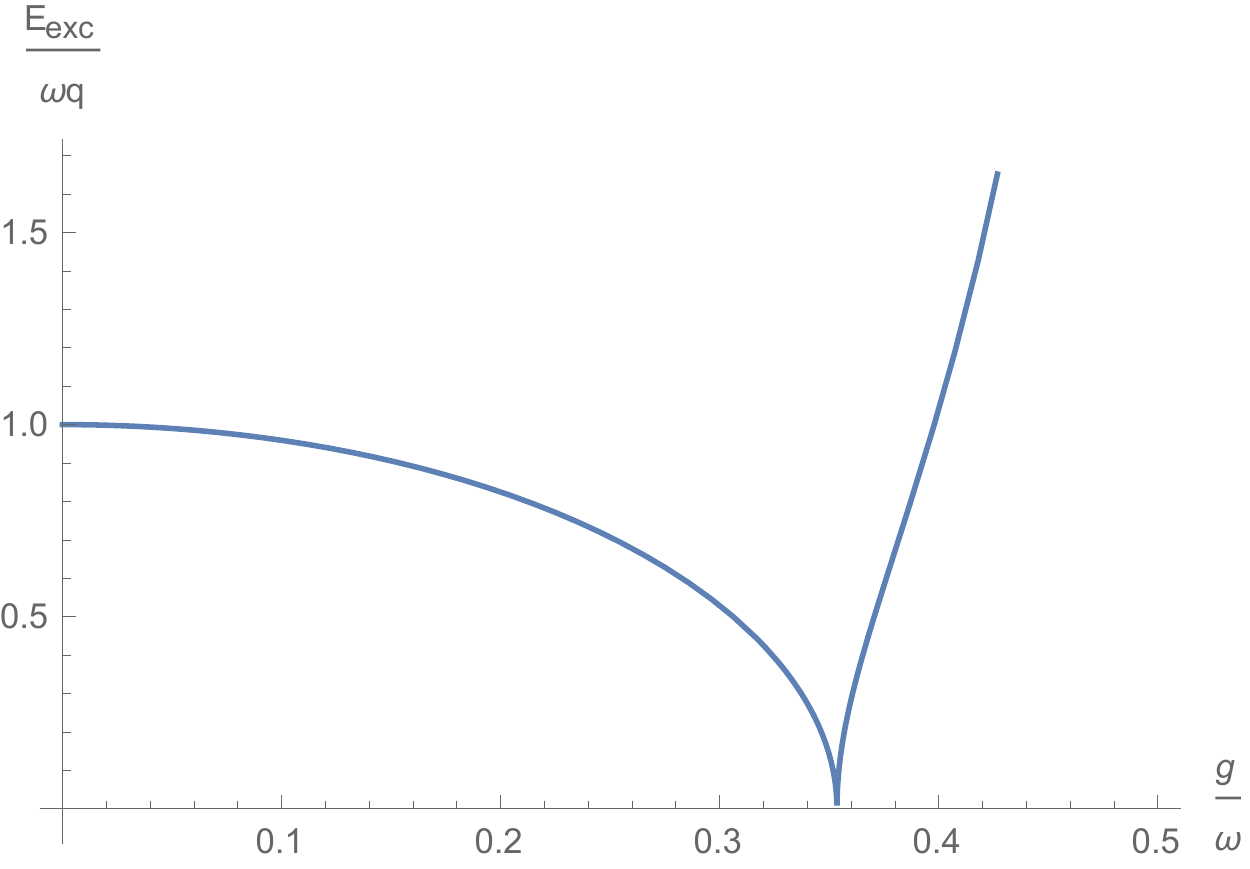}
\caption{Excitation energy $E_{exc}$ of the $\bop$ field, divided by $\omega_q$, for $\lambda=\frac{\omega}{2\omega_qN}=1$. In the regime of parameters considered, $\frac{E_{exc}}{\omega_q}$ is independent of the qubit number $N$. The cancellation of $E_{exc}$ for $g=g_t$ indicates the phase transition.}
\label{excitation energy}
\end{center}
\end{figure}

In the second phase, we also find squeezing properties for the $\bop$ ground state, with squeezing parameter:
\begin{widetext}
$$r_s^{(2)} = -\frac{1}{4}\ln\left(\frac{1+\frac{\alpha^2}{1-2\alpha^2}\left(3+\frac{\alpha^2}{1-\alpha^2}\right)-\frac{1-2\alpha^2}{1-\alpha^2-(16g^2/\omega^2)\alpha^2(1-\alpha^2)^2}}{1+\frac{\alpha^2}{1-2\alpha^2}}\right),$$ 
\end{widetext}
with $\alpha=\frac{\beta}{\sqrt{N}}$. The results for the $\aop$ field are identical to the mean-field case with a slight correction, i.e. we have a squeezed vacuum state with squeezing parameter $r_a^{(2)}$ which differs from $r_a^{(MF)}$ by a correction of order $\frac{1}{N}$ (see Appendix \ref{appendix second phase}).
Going back to the definition of the $\bop$ field (\ref{def Hosltein-Primakoff}), the behavior for the $\overrightarrow{J}$ operators can be summarized as follows:
in the first phase, $\overrightarrow{J}$ is polarized along the $z$-axis: $\meanvalue{\Jx}=\meanvalue{\Jy}=0$. When $g$ increases, the fluctuations of $\Jy$ are damped, while the fluctuations of $\Jx$ are amplified proportionally. The amplification factor goes to infinity when approaching the transition: even though the approximations we have used break down near the critical point, the divergence of the $\Jx$ fluctuations in our model may have observable consequences in an actual experiment. 
In the second phase, the $\overrightarrow{J}$ polarization will gradually evolve from the $z$-axis to the $x$-axis as $g$ increases. Near the spectral collapse $g=g_{collapse}=\frac{\omega}{2}$, $\overrightarrow{J}$ will be polarized along the $x$-axis, and we will have squeezing properties for the fluctuations in the $y$ and $z$ directions.
As of the $\aop$ field, we have a coherent vacuum state in the first phase; then, at the transition, the field acquires squeezing properties. In the second phase, the squeezing parameter increases with $g$ and diverges at the spectral collapse. The behavior of spin squeezing fluctuations is schematically depicted in Fig.\ref{Spin Bloch sphere}.
We can also characterize the ground state energy and the $\bop$ excitation energy in both phases: the results for the excitation energy are displayed on Fig.\ref{excitation energy}.

Finally, using the effective Hamiltonian in both phases, we can compute the critical exponents of several observables that exhibit critical behavior at the transition, that is, 
$$A(g \rightarrow g_t) \propto \lvert \frac{g-g_t}{g_t} \rvert^{\gamma_A},$$
$\gamma_A$ being the critical exponent of $A$. We can compare those results to the  critical exponents in the one-photon case \cite{Emary03,Vidal06}, with an interacting term of the form $\frac{g}{N}(\aop+\adag)(\Jplus+\Jminus)$ (note that this convention differs slightly from the one usually found in the literature, $\frac{g}{\sqrt{N}}(\aop+\adag)(\Jplus+\Jminus)$). 
In the one-photon model, the atomic and photonic excitations are hybridized to form polaritons. The energy spectrum exhibits two polaritonic branches, one of which goes to zero at the transition. In the limit $\omega_q \ll \omega$, however, the excitations cease to hybridize; the lowest- and highest-energy polaritons are purely atomic and photonic, respectively, which corresponds to our low-energy $\overrightarrow{J}$ excitations and high-energy $\aop$ ($\Ko$) excitations. Notice that this situation is similar to what is considered in Ref.\cite{Mottl12}, where an atomic ensemble interacts with a strongly detuned optical cavity. We compare the critical exponents of the one- and two-photon Dicke models in the regime $\omega_q \ll \omega$, in a such a way that the polaritonic branches have the same atomic and photonic components in both cases.    The results (computed for $\beta>0$) are summarized in Table  \ref{table critical exponents}.

\begin{table}[t]
\begin{center}
\begin{tabular}{| p{1cm} | p{2cm} | p{2cm} |}
\hline
 & Two-photon case & One-photon case \\\hline
$E_{exc}$ & $\frac{1}{2}$ & $\frac{1}{2}$ \\ \hline
$\Delta \hat{X}_d$ & $-\frac{1}{4}$ & $-\frac{1}{4}$\\ \hline
$\Delta \hat{X}_a$ & $0$ & $-\frac{1}{4}$ \\ \hline
\end{tabular}
\caption{Critical exponents of the excitation energy $E_{exc}$ and of the quadratures $\hat{X}_a=a+\adag$ and $\hat{X}_d=d+\opddag$. The one-photon interaction term is taken to be $\frac{g}{N}(\aop+\adag)(\Jplus+\Jminus)$. For the two-photon case, we took $\beta>0$. }
\label{table critical exponents}
\end{center}
\end{table}

Notice that the variance of $\hat{X}_a$ remains constant near the transition in the two-photon case, but diverges in the one-photon case, which is a marked difference. However, the diverging terms in the one-photon case are high-order terms; near the transition, $\Delta \hat{X}_a\sim 1 + O(\frac{1}{\sqrt{N}} \lvert \frac{g-g_t}{g_t} \rvert^{-\frac{1}{4}} )$. 
Thus, as long as one does not get too close to the transition, the behavior of $\hat{X}_a$ remains identical in the one-photon and the two-photon cases. When $\frac{g-g_t}{g_t}=O(N^{-2})$, though, the results begin to differ.  

On the other hand, the critical exponents are the same in both cases for $E_{exc}$ and $\hat{X}_d$.
Nevertheless, for these quantities, some of the higher-order terms we have neglected in our analysis may diverge faster near the transition that the terms we took into account, leading to the breakdown of our analysis when $g$ comes close enough to the critical value $g_t$. 
More precisely, the scaling parameter for $\Delta \hat{X}_d$ is no longer valid when $\frac{g-g_t}{g_t}$ becomes comparable with $\frac{1}{N}$. 


\section{Summary and Outlook}
\label{conclusions}
As a conclusion, we have shown the presence of a superradiant phase transition of the two-photon Dicke model in the ultrastrong coupling regime. We have characterized the behavior of the qubits and  bosonic field in both phases, and we have studied some of their critical properties near the transition. With respect to the one-photon case, fundamental differences arise in the behavior of the bosonic field $\aop$, which does not acquire macroscopic occupation in the second phase, and whose fluctuations do not diverge at the critical point, at least at the order considered. The pseudospin $\overrightarrow{J}$, on the other hand, does exhibit diverging fluctuations at the transition that could lead to observable phenomena which would mark the transition in experimentally accessible situations. Indeed, as an extension of this work, it would be interesting to analyze the spin-squeezing properties of the ground state of the system for a finite number of qubits. Quantum-phase transitions and the ground-state properties could be analyzed also modifying the symmetries of the model~\cite{Baksic14}, considering anisotropic couplings or including multiple bosonic fields. Finally, it would be interesting to study dynamical features and the emergence of quantum chaos~\cite{Emary03,Alvermann12,Bakemeier13} in the two-photon Dicke model. 


\section*{ACKNOWLEDGMENTS}
This work was supported by the French Agence Nationale de la Recherche (SemiQuantRoom project ANR-14-CE26-0029-01) and from University Sorbonne Paris Cite EQDOL contract.  S.F. acknowledges funding from the PRESTIGE program, under the Marie Curie Actions-COFUND of the FP7. E.S. and I.L.E. acknowledge funding from Spanish MINECO/FEDER FIS2015-69983-P, Basque Government IT986-16, UPV/EHU UFI 11/55.

\appendix

\section{Linear term expansion of the Hamiltonian}
\label{appendix linear expansion of H}
Let us consider an extension of our model, obtained by adding a linear term in $\aop$:
\begin{equation}
\hat{H}=\omega \adag \aop + \omega_q\Jz + \frac{g_2}{N}\Jx(\aop^2+\adagsq) + \frac{g_1}{N}\Jx(\aop+\adag).
\end{equation}
This gives, in the mean-field approximation:
\begin{subequations}
\begin{equation}
\Hop = \omega\adag \aop + \gtwoprime(\aop^2 + \adagsq) + \goneprime(\aop + \adag) + \omega_q\modbeta - \frac{\omega_qN}{2},
\end{equation}
\begin{equation}
\goneprime=\frac{g_1}{N}(\beta + \betaconj)\betasqr,
\end{equation}
\begin{equation}
\gtwoprime=\frac{g_2}{N}(\beta + \betaconj)\betasqr.
\end{equation}
\end{subequations}

We can shift the $\aop$ operator: $\aop = \cop - \alpha$. A proper choice of $\alpha$, namely, $\alpha=\frac{\goneprime}{\omega + 2\gtwoprime}$, allows to suppress the linear term in $\cop$. This gives us:

\begin{equation}
\Hop = \omega \cdag\cop + \gtwoprime(\cop^2 + \cdagsq) + \omega_q\modbeta - \frac{\omega_qN}{2} - \frac{(\goneprime)^2}{\omega + 2\gtwoprime},
\label{Heff c}
\end{equation}
which is just Hamiltonian (\ref{Heff a only}) with an additional constant term. Hence, $\aop$ in the ground state is in a displaced squeezed state, with squeezing factor $r_c =\frac{1}{2}\arctanh(\frac{2\gtwoprime}{\omega})$ and $\meanvalue{\aop}_{GS}=-\frac{\goneprime}{\omega + 2\gtwoprime}$. There is also a correction on the values of $\beta$ and $E_G$. Importantly, the presence of the term $\frac{(\goneprime)^2}{\omega + 2\gtwoprime}$ in (\ref{Heff c}), which is not invariant under the transformation $\beta \rightarrow - \beta$, lift the degeneracy between $\beta>0$ and $\beta<0$.

\section{Description of the method used to go beyond mean-field}
\label{appendix method Hwang description}
We are now going to detail the method used in \cite{article Hwang JC} and \cite{article Hwang Rabi}. We begin with (\ref{forme_generale_H}), and we assume that we have a two-dimensional pseudospin operator $\hat{\sigma}_z$. $\hat{H}_0$ is then a block diagonal operator with respect to the $\hat{\sigma}_z$ eigenbasis, and $\hat{H_c}$ is block anti-diagonal with respect to this eigenbasis (in the following, we will simply talk about block diagonal and block anti-diagonal operators, and about pseudospin eigenstates and eigenspaces: the reference to $\hat{\sigma}_z$ is implicit). We apply the unitary transformation $\Uop=e^{\Sop}$ to $\hat{H}$, which yields:

\begin{equation}
\Hop\sp{\prime}=e^{-\Sop} \Hop e^{\Sop}=\sum_{k=0}^{\infty}\frac{1}{k!}[\Hop,\Sop]^{(k)},
\label{eq_Hprime}
\end{equation}
with $[\Hop,\Sop]^{(k)}=[[\Hop,\Sop]^{(k-1)},\Sop]$ and $[\Hop,\Sop]^{(0)}=\Hop$. We impose $\Sop$ to be block anti-diagonal. If we split $\Hop\sp{\prime}$ into its block diagonal $\Hop\sp{\prime}_d$ and block anti-diagonal $\Hop\sp{\prime}_a$ parts, we get:

\begin{subequations}
\begin{equation}
\Hop\sp{\prime}_d=\sum_{k=0}^{\infty}\frac{1}{2k!}[\Hop_0,\Sop]^{(2k)}+\sum_{k=0}^{\infty}\frac{1}{2k+1!}[\epsilon \Hop_c,\Sop]^{(2k+1)},
\end{equation}
\begin{equation}
\Hop\sp{\prime}_a=\sum_{k=0}^{\infty}\frac{1}{2k+1!}[\Hop_0,S]^{(2k+1)} +\sum_{k=0}^{\infty}\frac{1}{2k!}[\epsilon \Hop_c,\Sop]^{(2k)}.
\end{equation}
\end{subequations}

These equations stem from the following relations: consider $D_1$ and $D_2$ two arbitrary block diagonal operators, and $O_1$ and $O_2$ two arbitrary block anti-diagonal operators. Then, $[D_1,D_2]$ and $[O_1,O_2]$ are block diagonal and $[O_1,D_1]$ is block anti-diagonal.
 
At this point, the idea is to expand $\hat{S}$ with respect to $\epsilon$: $\Sop=\epsilon \Sop_1 + \epsilon^2 \Sop_2 +...$, with $\Sop_i$ block anti diagonal for all $i$. By properly choosing the $\Sop_i$ for $i$ from $1$ to $p$ with $p$ arbitrary, it is possible to cancel $\Hop\sp{\prime}_a$ up to order $\epsilon^p$, thus decoupling the pseudospin eigenspaces up to this order.

Our case, however, is slightly different, because we have an infinite number of eigenstates instead of only two. When SU(1,1) is presented in terms of quadratic combinations of creation/annihilation operators the relevant representations are those with Bargmann parameter $q=\frac{1}{4}$ and $q=\frac{3}{4}$. The first one corresponds to even number of $\hat{b}$ excitations, while the second one is associated with an odd number \cite{article Duan Bargmann representation and Rabi analytic solution}. They are not connected by any of the operators of the algebra. Since we will focus on the ground state, we concentrate on the $q=\frac{1}{4}$ case. Then, we have an infinity of eigenstates $\ket{n}_K$, with $\Ko\ket{n}_K=(n+\frac{1}{4})\ket{n}_K$. This means that we have to modify slightly the method above if we want to apply it to bigger Hilbert spaces.

For this, let us consider the operators diagonal in the $\ket{n}_K$ basis, which can be written as $\sum_{n=0}^{\infty}\alpha^0_n\ket{n}_K\bra{n}_K$ with arbitrary $\alpha^0_n$. Let us call the ensemble of these operators $M_{(0)}$; $\hat{K}_0$, for instance, belongs to this ensemble. Then, let us call $M_{(1)}$ the ensemble of all operators of the form $\sum_{n=0}^{\infty}\alpha^1_n\ket{n}_K\bra{n+1}_K + \sum_{n=1}^{\infty}\beta^1_n\ket{n}_K\bra{n-1}_K$, with arbitrary coefficients; $\Kp + \Km$ is an element of $M_{(1)}$. $M_{(2)}$ contains the operators of the form $\sum_{n=0}^{\infty}\alpha^2_n\ket{n}_K\bra{n+2}_K + \sum_{n=2}^{\infty}\beta^2_n\ket{n}_K\bra{n-2}_K + \sum_{n=0}^{\infty}\gamma^2_n\ket{n}_K\bra{n}_K$; and in a general way, we define the ensemble $M_{(j)}$ that contains operators that can be written as:

\begin{subequations}
\begin{equation}
\hat{F}(j) + \hat{F}(j-2) + \hat{F}(j-4)+...=\sum_{p<\frac{j}{2}}\hat{F}(j-2p),
\end{equation}
\begin{equation}
\hat{F}(i)=\sum_{n=0}^{\infty}\rho^i_n\ket{n}_K\bra{n+i}_K + \sum_{n=i}^{\infty}\chi^i_n\ket{n}_K\bra{n-i}_K. 
\end{equation}
\end{subequations}

Let us note here that there is a redundancy in this definition: an element of $M_{(j)}$ also belongs to $M_{(j+2)}$, for all $j$. We will use the following property: for all $\hat{A}$ belonging to $M_{(i)}$ and $\hat{B}$ belonging to $M_{(j)}$, $\hat{C}=[\hat{A},\hat{B}]$ belongs to $M_{(i+j)}$, which we denote symbolically by 
\begin{equation}
[M_{(i)},M_{(j)}]_{op} \subseteq M_{(i+j)}.
\label{comm_rel_Mi}
\end{equation}

The idea in the following will be to isolate the elements of the various $M_{(i)}$ and to cancel those that couple the eigenstates of $\Ko$ to a certain order.

\section{Study of the first phase}
\label{appendix first phase}
As indicated in the main text, we perform the Holstein-Primakoff transformation, as well as the transformation (\ref{def_Ko}-\ref{def_Km}), and we consider the case $\beta=0$: $\bop=\beta+\dop=\dop$. This gives us for the Hamiltonian:
\begin{equation}
\Hop = -\frac{1}{2}\omega -\frac{\omega_q N}{2} + 2\omega \Ko + \omega_q \opddag \dop + \frac{2g}{\sqrt{N}}(\Kp + \Km)(\dop+\opddag). 
\end{equation}
With a redefinition of parameters $\omega_1=\frac{\omega_q N}{2\omega}=O(1)$ and $\omega_2=\frac{g}{\omega}=O(1)$, we get:

\begin{eqnarray}
\hat{H}_1&=&\frac{(\Hop+\frac{1}{2}\omega +\frac{\omega_q N}{2})}{2\omega}\nonumber \\
&=&\Ko + \frac{\omega_1}{N}\opddag \dop + \frac{\omega_2}{\sqrt{N}}(\dop+\opddag)(\Kp + \Km),
\label{eqH1}
\end{eqnarray}
which is explicitly of the form \eqref{forme_generale_H}; $\hat{H}_0=\Ko + \frac{\omega_1}{N}\opddag \dop$ does not couple the $\Ko$ eigenstates, but $\hat{H}_c=\omega_2(\dop+\opddag)(\Kp + \Km)$ does. In our case, the small parameter is $\frac{1}{\sqrt{N}}$.\\

The generalized method goes as follows: all the operators implied in our calculations can be (not uniquely) decomposed into elements belonging to the different $M_{(i)}$: $\hat{A}=\sum_{i=0}^{n_{max}} \hat{A}^{(i)}$, where $n_{max}$ can take an arbitrary high value. We then propose the following decompositions:  
\begin{subequations}
\begin{equation}
\Hop_1 = \Hop_1^{(0)} + \Hop_1^{(1)},
\end{equation}
\begin{equation}
\Hop_1^{(0)}=\Ko + \frac{\omega_1}{N}\opddag \dop,
\end{equation}
\begin{equation}
\Hop_1^{(1)}=\frac{\omega_2}{\sqrt{N}}(\dop+\opddag)(\Kp + \Km)=\frac{1}{\sqrt{N}}\hat{V}^{(1)},
\end{equation}
\end{subequations}

\begin{equation}
\Sop=\frac{1}{\sqrt{N}}\Pop + \frac{1}{N}\Qop + \frac{1}{N}\Rop + O\left(\frac{1}{N\sqrt{N}}\right),
\end{equation}
with $\Pop$ and $\Qop$ that belong to $M_{(1)}$ and $\Rop$ that belongs to $M_{(2)}$. Using \eqref{comm_rel_Mi} and \eqref{eq_Hprime}, we then obtain the following decomposition for $\Hop_1\sp{\prime}=e^{-\hat{S}}\Hop_1e^{\hat{S}}$:
\begin{subequations}
\begin{equation}
\Hop_1\sp{\prime(0)}=\Hop_1^{(0)}=\Ko + \frac{\omega_1}{N}\opddag \dop,
\end{equation}
\begin{equation}
\Hop_1\sp{\prime(1)}=\frac{1}{\sqrt{N}}\hat{V}^{(1)} + \frac{1}{\sqrt{N}}[\Ko,\Pop] + \frac{1}{N}[\Ko,\Qop] + O\left(\frac{1}{N\sqrt{N}}\right),
\end{equation}
\begin{equation}
\Hop_1\sp{\prime(2)}=\frac{1}{N}[\hat{V}^{(1)},\Pop] + \frac{1}{2N}[[\Ko,\Pop],\Pop] + \frac{1}{N}[\Ko,\Rop] + O\left(\frac{1}{N\sqrt{N}}\right),
\end{equation}
\begin{equation}
\Hop_1\sp{\prime(3)}=O\left(\frac{1}{N\sqrt{N}}\right),
\end{equation}
\end{subequations}
since $\Hop_1\sp{\prime(1)}$ couples the eigenvalues of $\Kop$, we are going to cancel it at order $\frac{1}{N}$. For this purpose, we need $\hat{V}^{(1)} + [\Ko,\Pop]=0$ and $[\Ko,\Qop]=0$. We note that these conditions can be met by the following choice of operators:
\begin{subequations}
\begin{equation}
\Pop=-\omega_2(\dop+\opddag)(\Kp-\Km),
\end{equation}
\begin{equation}
\Qop=0.
\end{equation}
\end{subequations}
As of the $\Hop_1\sp{\prime(2)}$ term, we find, using the above expressions for $\Pop$ and $\Qop$:
\begin{equation}
\Hop_1\sp{\prime(2)} = -\frac{4\omega_2^2}{2N}(\dop + \opddag)^2\Ko + \frac{1}{N}[\Ko,\Rop] + O\left(\frac{1}{N\sqrt{N}}\right).
\end{equation}
In general, we should choose $\Rop$ so as to cancel the non-diagonal terms of this operator. Here, though, $-\frac{4\omega_2^2}{2N}(\dop + \opddag)^2\Ko$ is already diagonal; it is thus sufficient to set $\Rop=0$. We speculate that, for higher order expansions, if $\Hop$ can be made diagonal at order $\left(\frac{1}{\sqrt{N}}\right)^{i-1}$, we can make it diagonal at order $\left(\frac{1}{\sqrt{N}}\right)^{i}$ by adding terms of the form $\left(\frac{1}{\sqrt{N}}\right)^{i}\hat{T}_p$ to the expansion of $\hat{S}$, with $\hat{T}_p$ that belongs to $M_{(p)}$ and $p \le i$. It could be interesting in a future study to consider higher-order terms in a more systematic way to confirm or to invalidate this conjecture. 

In the end, all those manipulations yield:
\begin{equation}
\hat{H}_1\sp{\prime}=e^{-\Sop} \Hop_1 e^{\Sop}=\Ko + \frac{\omega_1}{N}\opddag \dop - \frac{2\omega_2^2}{N}(\dop+\opddag)^2\Ko + O\left(\frac{1}{N\sqrt{N}}\right).
\label{H transformé 1ère phase}
\end{equation}
This Hamiltonian does commute with $\Ko$; projection in the $\Ko$ ground state gives $\Ko\rightarrow\frac{1}{4}$ (according to the definition \eqref{def_Ko}). Restoring the constants in the Hamiltonian yields the operator described in the main text:
\begin{equation}
\Hop = -\frac{\omega_qN}{2} + \omega_q \opddag \dop - \frac{g^2}{N\omega}(\dop+\opddag)^2 + O\left(\frac{\omega}{N\sqrt{N}}\right).
\label{Heff_1erephase}
\end{equation}
This effective Hamiltonian describes the $\bop$ fluctuations near the ground state. Being quadratic in $\dop$, it can be diagonalized by using a Bogoliubov transformation of parameter $r_s^{(1)}=\frac{1}{4}\ln(1-\frac{4g^2}{N\omega\omega_q})<0$, which corresponds to a squeezed vacuum state as described in the main text. The state of $\aop$ is given by the ground state of $\Ko$; here it is just the coherent vacuum state, as in the mean-field scenario. Finally, the ground state energy and the $\dop$ excitation energy are computed: we obtain
\begin{subequations}
\begin{equation}
E_{exc}^{(1)}=\omega_q\sqrt{1-\frac{4g^2}{N\omega\omega_q}}, 
\end{equation}
\begin{equation}
E_G^{(1)}=-\frac{\omega_qN}{2} + \frac{E_{exc}^{(1)}-\omega_q}{2}. 
\end{equation}
\label{EG_1}
\end{subequations}

\section{Study of the second phase}
\label{appendix second phase}
We proceed in a similar fashion, using HP transformation once more, and setting $\bop=\beta + \dop$; but this time $\beta\ne 0$. As a starting point, we use the value of $\beta$ obtained by our mean-field analysis. For ease of notation, we define the following parameters:\\

\begin{subequations}
\begin{equation}
\alpha=\frac{\beta}{\sqrt{N}}=O(1),
\end{equation}
\begin{equation}
\chi=\sqrt{1-\frac{\beta^2}{N}}=\sqrt{1-\alpha^2}=O(1),
\end{equation}
\begin{equation}
\delta=1-\frac{\beta^2}{N-\beta^2}=O(1).
\end{equation}
\end{subequations}

Expanding the Hamiltonian gives:
\begin{widetext}
\begin{eqnarray}
\vspace{15pt}
\Hop_2 & = & (\Hop+\frac{1}{2}\omega + \frac{\omega_qN}{2} -\omega_qN\alpha^2)/2\omega \nonumber\\
\vspace{15pt}
 & = & \lambda_0 \Kop + \frac{\lambda_1}{\sqrt{N}}(d+\opddag) + \frac{\lambda_2}{\sqrt{N}}(d+\opddag)(\Kpp + \Kmp) + \frac{\lambda_3}{\sqrt{N}}(d+\opddag)\Kop \\ 
\vspace{15pt}
 & & + \frac{\lambda_4}{N}\opddag d - \frac{2}{N}\Kop \hat{V}_1(\dop) + \frac{1}{N}(\Kpp + \Kmp) \hat{V}_2(\dop) + O\left(\frac{1}{N\sqrt{N}}\right). \nonumber
\end{eqnarray}
\end{widetext}

Where we have defined new operators and parameters:
\begin{subequations}
\begin{equation}
\Kop =\cosh(2r_a^{(2)})\Ko + \frac{1}{2}\sinh(2r_a^{(2)})(\Kp + \Km),
\end{equation}
\begin{equation}
\Kpp + \Kmp = \cosh(2r_a^{(2)})(\Kp + \Km)+ 2\sinh(2r_a^{(2)})\Ko,
\end{equation}
\begin{eqnarray}
r_a^{(2)} = & \frac{1}{2}\arctanh\left(\frac{4g\alpha\chi}{\omega} + \frac{g\alpha}{\omega\chi N}\right) \nonumber\\ 
 & = \frac{1}{2}\arctanh\left(\frac{2\gprime}{\omega} + \frac{g\alpha}{\omega\chi N}\right), 
\label{def ra2}
\end{eqnarray}
\label{Bogo K0}
\end{subequations}

\begin{subequations}
\begin{equation}
\lambda_0=\cosh(2r_a^{(2)}) - \left(\frac{4g\alpha\chi}{\omega} + \frac{g\alpha}{\omega\chi N}\right)\sinh(2r_a^{(2)}),
\end{equation}
\begin{equation}
\lambda_1=\frac{\omega_qN\alpha}{2\omega},
\end{equation}
\begin{equation}
\lambda_2 = \frac{g\chi\delta}{\omega}\cosh(2r_a^{(2)}),
\end{equation}
\begin{equation}
\lambda_3=-2\sinh(2r_a^{(2)})\frac{g\chi\delta}{\omega},
\end{equation}
\begin{equation}
\lambda_4=\frac{\omega_qN}{2\omega},
\end{equation}
\begin{equation}
\hat{V}_1(\dop)= \sinh(2r_a^{(2)}) \left(-\frac{g\alpha}{\chi \omega}\opddag \dop - \frac{g}{\omega} \left(\frac{\alpha}{2\chi} + \frac{\alpha^3}{4\chi^3}\right) (\dop+\opddag)^2\right),
\end{equation}
\begin{equation}
\hat{V}_2(\dop)= \cosh(2r_a^{(2)}) \left(-\frac{g\alpha}{\chi \omega}\opddag \dop - \frac{g}{\omega} \left(\frac{\alpha}{2\chi} + \frac{\alpha^3}{4\chi^3}\right)(\dop+\opddag)^2\right).
\end{equation}
\end{subequations}

Let us note that the definition (\ref{Bogo K0}) amounts to a Bogoliubov transformation of parameter $r_a^{(2)}$ for the $\aop$ field. We seek to decouple the eigenspaces of $\Kop$; for this, we apply a transformation $e^{-\Sop} \Hop_2 e^{\Sop}$ with $\Sop=\frac{1}{\sqrt{N}}\Sop_1 + \frac{1}{N}\Sop_2$. Using the procedure described earlier, we propose the following operators:

\begin{subequations}

\begin{equation}
\Sop_1=-\frac{\lambda_2}{\lambda_0}(\dop+\opddag)(\Kpp - \Kmp),
\end{equation}
\begin{equation}
\Sop_2=(\Kpp - \Kmp) \left(\frac{\lambda_3\lambda_2}{\lambda_0^2}(\dop+\opddag)^2 - \frac{\hat{V}_2(\dop)}{\lambda_0}\right).
\end{equation}

\end{subequations}

This gives us an Hamiltonian that commutes with $\Kop$:
\begin{widetext}
\begin{equation}
e^{-\Sop} \Hop_2 e^{\Sop} = \lambda_0\Kop + \frac{\lambda_3}{\sqrt{N}}(\dop+\opddag)\Kop + \frac{\lambda_1}{\sqrt{N}}(\dop+\opddag) + \frac{\lambda_4}{N}\opddag \dop - \frac{2}{N}\Kop \hat{V}_1(\dop) - 4\frac{\lambda_2^2}{2N\lambda_0}(\dop+\opddag)^2\Kop + O\left(\frac{1}{N\sqrt{N}}\right).
\end{equation}
\end{widetext}

As previously, projection in the ground state of $\Kop$ yields an effective Hamiltonian describing the fluctuations of $\dop$ above the ground state. Note that in contrast to the first phase, $e^{-\Sop} \Hop_2 e^{\Sop}$ contains a term linear in $\dop + \opddag$. $e^{-\Sop} \Hop_2 e^{\Sop}$ can be seen as an effective potential for the $\dop$ (or $\bop$) field; adding this linear term shifts its minimum, which amounts to change the value of $\beta$. In our case, the term adds a correction of order $\frac{1}{\sqrt{N}}$ to the value of $\beta$ (keeping in mind that the mean-field value of $\beta$ is of order $\sqrt{N}$). We can thus absorb the linear term by adding this correction to $\beta$; once it is done, $e^{-\Sop} \Hop_2 e^{\Sop}$ becomes quadratic in $\dop$ and can be diagonalized by a Bogoliubov transformation of parameter $r_s^{(2)}$, which gives squeezing properties for $\bop$. Going back to the $\overrightarrow{J}$ operators, we have:\\

\vspace{-20pt}

\begin{subequations}
\begin{equation}
\Jx = \beta\sqrt{N-\beta^2} + \frac{N-2\beta^2}{2\sqrt{N-\beta^2}}\hat{X}_d + (d^2 \hspace{5pt} term),
\end{equation}
\begin{equation}
\Jy = - \frac{\sqrt{N-\beta^2}}{2}\hat{P}_d + (d^2 \hspace{5pt} term),
\end{equation}
\begin{equation}
\Jz = (\beta^2-\frac{N}{2}) + \beta \hat{X}_d + (d^2 \hspace{5pt} term),
\end{equation}
\end{subequations}
with $\hat{X}_d$ et $\hat{P}_d$ the quadratures of the $\dop$ field. Using the squeezing properties of $\dop$ in both phases, we retrieve the properties described in the main text and in Fig.\ref{Spin Bloch sphere}. As of the state of $\aop$, it corresponds to the ground state of $\Kop$; according to the definition (\ref{Bogo K0}), this is just a squeezed vacuum state with squeezing parameter $r_a^{(2)}$. This result is identical to what we obtained in the mean-field scenario (\ref{def ra mean-field}), up to a small correction in the definition of $r_a^{(2)}$: $r_a^{(2)}-r_a^{(MF)}=\frac{1}{2}\arctanh\left(\frac{2\gprime^{(MF)}}{\omega} + O\left(\frac{1}{N}\right)\right) - \frac{1}{2}\arctanh\left(\frac{2\gprime^{(MF)}}{\omega}\right) = O(\frac{1}{N})$ (while $r_a^{(MF)}$ is of order O(1)).

Finally, the computation of the  $\dop$ excitation energy gives:
\begin{widetext}
\begin{equation}
E_{exc}^{(2)}=\omega_q\sqrt{\left(1+\frac{\alpha^2}{1-2\alpha^2}\right) \left(1+\frac{\alpha^2}{1-2\alpha^2}\left(3+\frac{\alpha^2}{1-\alpha^2}\right)-\frac{1-2\alpha^2}{1-\alpha^2-(16g^2/\omega^2)\alpha^2(1-\alpha^2)^2}\right)}.
\label{EG_2}
\end{equation}
For the sake of simplicity, we do not show here the expression of the  ground state energy $E_G^{(2)}$ for the second phase. The plot on Fig.\ref{excitation energy} is obtained by combining $E_{exc}^{(1)}$ and $E_{exc}^{(2)}$. 
\end{widetext}


\begin{thebibliography}{20}
\bibitem{note subspace}
Since $H$ and $\overrightarrow{J}^2$ commute, we are allowed to consider each angular momentum subspace independently in our quest for the Hamiltonian ground state. We also expect the ground state of $\Hop$ to belong to the $\mathcal{H}_{\frac{N}{2}}$, meaning that only this subspace is relevant for our study.
This assumption is consistent with \cite{Emary03} and \cite{Baksic14} and can be justified as it follows. For vanishing value of the coupling constant $g$, the ground state is doubly degenerate, and the qubits in the ground state are all polarized along the z axis. The spin state belongs to the space spanned by $\ket{\uparrow_z\uparrow_z\uparrow_z..}$ and $\ket{\downarrow_z\downarrow_z\downarrow_z..}$; hence, the ground state belongs to $\mathcal{H}_{\frac{N}{2}}$ for $g\rightarrow 0$. When $g$ increases, the energy levels are modified by the presence of the interaction term. The states associated with high absolute values of $\hat{J}_x$ will have a strong correction, due to the strong coupling between the qubits and the field. Due to this, the lowest energy eigenstate of $\mathcal{H}_{\frac{N}{2}}$ will be lowered faster than any other state, and thus will remain the lowest energy eigenstate overall. As a consequence, the ground state of $\Hop$ should belong to $\mathcal{H}_{\frac{N}{2}}$ for all values of $g$.
\bibitem{note zero omegaq}
We can also note that the limiting case for this parameter regime is $\omega_q=0$. In that case, the two degenerate ground states are $\ket{\uparrow_x\uparrow_x\uparrow_x..;\xi(r)} \pm \ket{\downarrow_x\downarrow_x\downarrow_x..\xi(-r)}$, where $\xi(r)$ means that the bosonic field is in a squeezed vacuum with squeezing parameter $r = \frac{1}{2}\arctanh(\frac{2g}{\omega})$; these results coincide with the behavior of the full model near the collapse point $g=\frac{\omega}{2}$.
\end{thebibliography}
\end{document}